\documentclass[prb,twocolumn,amsmath,amssymb]{revtex4-2}

\usepackage{graphicx}
\usepackage{dcolumn}
\usepackage{bm}
\usepackage{mathrsfs}
\usepackage{natbib}
\usepackage{amsmath}
\usepackage{amssymb}
\usepackage{Jonasmacros}
\usepackage{mathptmx}
\usepackage{pifont}
\usepackage{times}
\usepackage{txfonts}

\usepackage[colorlinks,citecolor=magenta,linkcolor=red]{hyperref}
\usepackage[usenames,dvipsnames,svgnames]{xcolor}

\date{\today} 
\begin{document}

\title{Breaking of Time-Reversal Symmetry and Onsager Reciprocity in Chiral Molecule Interfacd with an Environment}

\author{Jonas Fransson}
\email{Jonas.Fransson@physics.uu.se}
\affiliation{Department of Physics and Astronomy, Box 516, 751 21, Uppsala University, Uppsala, Sweden}

\begin{abstract}
Molecular closed shell structures are known to form spin-singlet configurations, resulting from the spin-exchange associated with electron-electron interactions. While the vanishing total spin-moment is an immanent property of the spin-singlet, the vanishing local moments is a result of fluctuations between degenerate spin-configurations. Here, it is demonstrated that the spin-configuration of a chiral molecule is enantiospecifically locked when coupled to an electron reservoir. The coupling opens up the molecular closed shell structure, which  broadens the energy levels. Together with the molecular spin-orbit coupling, this dissipative coupling generates an effective spin-splitting of the molecular energy levels and facilitates stabilization of a chirality determined spin-configuration. Simultaneously, the charge molecular charge distribution is shown to depend linearly on the magnetization of the reservoir such that no linear response regime is established with respect to the external magnetization. Accordingly, the Onsager reciprocity theorem does not apply to a chiral molecule attached to a reservoir, hence, providing a theoretical foundation for the chirality induced spin selectivity effect.
\end{abstract}

\maketitle

\section{Introduction}
The chirality induced spin selectivity effect \cite{Science.283.814,Science.331.894,NanoLett.11.4652,ChemRev.124.1950} challenges many aspects that are held to be true in with respect to, e.g., magnetism, current response, and finite structures. Questions about time-reversal symmetry and Onsager reciprocity \cite{NanoLett.19.5253,PhysRevB.99.024418,PhysRevB.102.035445,ACSNano.16.4989,ACSNano.18.6028}, and whether spin-signatures can be acquired in two-terminal measurements \cite{PhysRevB.93.075407,NanoLett.20.6148}, are part of an intensive debate on what a comprehensive theory should comprise.

Of special interest are the transport measurements since the experimental observations display many seemingly contradicting components. In the context of charge transport, the chirality induced spin selectivity effect is a measure of an asymmetric current response to the external magnetic conditions \cite{NanoLett.11.4652,IsraelJChem.62.e202200046}. However, for the Onsager reciprocity to hold true, time-reversal symmetry must remain intact in equilibrium \cite{PhysRev.37.405}. By contrast, the experimental observations strongly indicate that the chirality induced spin selectivity effect varies only weakly in a range of voltage biases up to a few volts around equilibrium \cite{NanoLett.11.4652,JPhysChemC.124.10776,NanoscaleHoriz.8.320,Small.20.2308233,JPhysChemLett.15.6605,ACSNano.19.17941}. This issue would be resolved should the chiral molecule acquire a frozen spin-configuration when interfaced with an electron reservoir, something which necessitates a time-reversal symmetry breaking mechanism.

The possibility of a spontaneously time-reversal broken state was discussed in Ref. \cite{PhysRevB.108.235154} in terms of chiral atomic structures interacting with a reciprocal environment. The important result for the present context is that the time-evolution of the spin vector inevitably leads to one specific solution and that this solution is determined by chirality. An analogous result, based on a correlated molecular structure being interfaced with a metallic reservoir, was discussed in \cite{JPhysChemLett.14.5119}.

The purpose of this article is to demonstrate that chiral molecules can acquire a stable non-fluctuating spin-configuration when interfaced with a reciprocal environment, e.g., an electron reservoir which can be achieved using a metallic or semi-conducting substrate. The basic observations pertaining to molecules are ({\it i}) that the spin-singlet state of a closed shell structure is composed of correlated electron states \cite{Correlated}, ({\it ii}) the spin-orbit coupling provides a connection between the molecular structure and the electron spin, ({\it iii}) that the coupling to the electron reservoir provides a dissipative channel which, together with the spin-orbit coupling, leads to an effective local Zeeman-like spin-splitting of the energy levels, and ({\it iv}) the electronic structure depends linearly on the magnetization of the electron reservoir.

The first observation, that the spin-singlet state is a composition of correlated electron states, is crucial for how the electronic structure should be perceived. Contrary to the notion that may hold for semi-conductors and some metals, in which the itinerant electrons can be successfully treated as independent particles, the valence of a closed shell molecule is always a correlated configuration. This dichotomy effectively means that while the spin expectation value of itinerant electrons is independent of the spin of other electrons and, therefore, averages to zero, the spin expectation value of a valence electron in a molecule is not independent of the spins of other electrons. Hence, the molecule momentarily assumes a specific spin-configuration which is switched, e.g., by thermal fluctuations, into another in the next moment, such that the spin expectation value summed over the configurations vanishes. However, here it will be shown that the molecular spin configuration is locked into a specific configuration when the molecule is interfaced with an environment. The acquired configuration is, furthermore, determined by the molecular chirality.

The second observation, concerning the spin-orbit coupling, is crucial for the spin-configuration to lock into a specific state. This conclusion may at first seem rather counterintuitive, keeping in mind that spin-orbit coupling mixes the spins of the electrons such that the spin is no longer a good quantum number. Nevertheless, in the present context, the spin-orbit coupling is the only viable connection between the molecular structure and the electronic spin and despite that it cannot in itself lead to a spin-configuration, when combined with dissipative processes it contributes to driving the system into a spontaneously time-reversal broken state.

Therefore, the third observation, regarding dissipation, is indispensable for the formation of a local spin-configuration in chiral molecules. By connecting the molecule to an electron reservoir, the closed shell structure opens up enabling exchange of particles between the two subsystems. By this particle exchange, the molecular states are no longer eigenstates of the system while the vast number of choices offered by the reservoir leads to that the life-times of the molecular states become finite. A finite life-time is equivalent with dissipative processes as energy is lost to the reservoir in the decay.

The fourth observation, finally, indicates that the electronic structure depends linearly on the magnetization of the electron reservoir, henceforth referred to as the external magnetization. Let $\bfE_m=\dote{m}\sigma^0+\bfepsilon_m\cdot\bfsigma$ represent the effective energy spectrum at the molecular site $m$, where $\dote{m}$ and $\bfepsilon_m$ denote the spin-independent and spin-dependent portions, and where $\sigma^0$ and $\bfsigma$ are the $2\times2$-unit matrix and vector of Pauli matrices, respectively. In the following it is shown that both $\dote{m}$ and $\bfepsilon_m$ depend linearly on the external magnetization. For both para- and diamagnetic matter, the latter portion $\bfepsilon_m$ is expected to vary linearly with the external magnetization. By contrast, linear dependence of the former portion, $\dote{m}$, on the external magnetization implies that the spectrum shifts asymmetrically as function of the external magnetization. Consequently, these spectral variations implies that one cannot define a linear response regime with respect to the external magnetization for the composite structure, implying that the Onsager reciprocity relations are inapplicable in this context.

Chirality has an exclusive role in this discussion since it introduces a directionality in the set-up which can be understood as an anisotropy. The curvature associated with the chiral distribution of the nuclei provides an anisotropic environment for the spin-moments at each nucleus and, thereby, facilitates an ordered configuration of the spins. Hence, planar achiral molecules of zig-zag type also acquire a spin-distribution, however, only transverse to the plane defined by the molecule. Therefore, there is no associated magneto-resistance of the type that is possible with chiral molecules.



\section{Results}
\subsection{Modeling the correlated molecular state}
\begin{figure}[t]
\begin{center}
\includegraphics[width=\columnwidth]{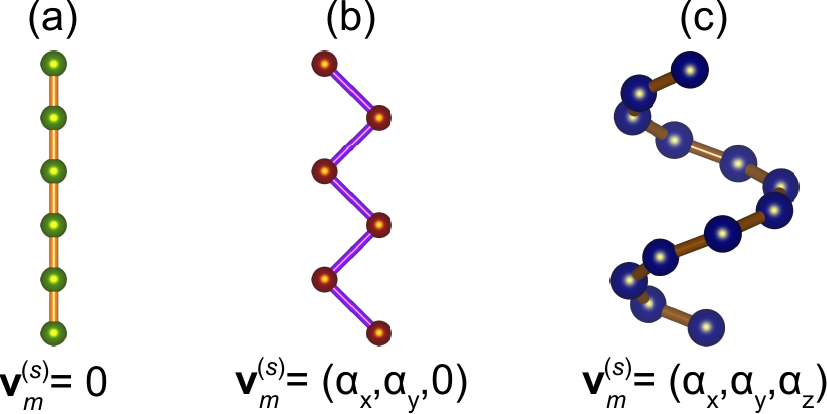}
\end{center}
\caption{Curvature vector $\bfv_m^{(s)}$ for three geometries, (a) linear chain, (b) zig-zag chain, and (c) helical chain.}
\label{fig-schematic}
\end{figure}

The three observations discussed above are tangible in a Hamiltonian model, $\Hamil_\text{mol}$, for correlated electrons in a molecular structure. Consider a molecular chain comprising $\mathbb{M}$ sites with a single electron energy level $\dote{m}$ per site, $m=1,2,\ldots,\mathbb{M}$, connected via nearest and next-nearest neighbor hopping $t_\nu$ and $\lambda_\nu$, respectively, where $\nu=0$ refers to elastic processes and $\nu=1$ to vibrationally assisted processes. Effectively \cite{PhysRevB.102.235416}
\begin{subequations}
\label{eq-mol}
\begin{align}
\Hamil_\text{mol}=&
	\sum_{mn}
		\psi^\dagger_m
		\left(
			E-H_0+H_1\sum_iQ_i
		\right)_{mn}
		\psi_n
		+
		\Hamil_\text{vib}
	,
\\
\Bigl(
	H_\nu
	\Bigr)_{mn}=&
	\sum_{s=\pm1}
		\Bigl(
			-t_\nu\delta_{nm+s}
			+
			i\lambda_\nu\bfv_m^{(s)}\cdot\bfsigma\delta_{nm+2s}
		\Bigr)
	,
\end{align}
\end{subequations}
where the spinor $\psi_m=(\psi_{m\up}\ \psi_{m\down})^t$ and $E=\diag{\dote{m}}{m}$ is a diagonal matrix. The term proportional to $\lambda_\nu$ provides spin-orbit coupling, where the vector $\bfv_m^{(s)}$ defines the curvature between the sites $m$, $m+s$, and $m+2s$ through the vector product $\bfv_m^{(s)}=\hat{\bfd}_{m,s}\times\hat{\bfd}_{m+s,s}$. Here, $\bfd_{m,s}=\bfr_m-\bfr_{m+s}$, $\hat{\bfd}_{m,s}=\bfd_{m,s}/|\bfd_{m,s}|$, denotes the distance between the two sites at $\bfr_m$ and $\bfr_{m+s}$. In this construction $\bfv_{m+2s}^{(\bar{s})}=-\bfv_m^{(s)}$, where $\bar{s}=-s$. Three types of structures and corresponding curvature vectors are illustrated in Fig. \ref{fig-schematic}. Particularly, it is noticeable that the zig-zag chain (b) is non-trivial, such that there may be resulting spin-mixing. In the chiral (c) structure, by contrast, the curvature also opens up for a longitudinal component of the spin-orbit coupling, which is crucial in the present context as will be seen below.

The vibrational part $\Hamil_\text{vib}=\sum_i\omega_ib^\dagger_ib_i+\sum_{ijk}\Phi_{ijk}Q_iQ_jQ_k$ describes the molecular vibrations with harmonic energies $\omega_i$ and anharmonic corrections with coupling $\Phi_{ijk}$, in terms of the bosonic operators $b_i$, $b^\dagger_i$, and $Q_i=b_i+b^\dagger_i$. The molecular vibrations and electrons are coupled through $H_1\sum_iQ_i$, and since it is the coherent molecular vibrations that couple to the electron spin \cite{PhysRevB.102.235416}, it is assumed that all vibrational modes couple to all electrons in the molecule.

The molecule is mounted between a left ($L$) and right ($R$) lead, modelled as free electron reservoirs $\Hamil_\chi=\sum_{\bfk\in\chi}\psi^\dagger_\bfk\bfepsilon_\bfk\psi_\bfk$, $\chi=L,R$, where $\psi_\bfk=(\psi_{\bfk\up}\ \psi_{\bfk\down})^t$ is the spinor for electrons in the reservoir at the energy $\dote{\bfk}$. The reservoir is assumed to be in local equilibrium such that the Fermi function $f_\chi(\omega)=f(\omega-\mu_\chi)$ is well-defined at the chemical potential $\mu_\chi$. The molecule and reservoirs are coupled through $\Hamil_T=\sum_{\bfk\in L}\psi^\dagger_{\bfk}\bfu_{\bfk1}\psi_1+\sum_{\bfk\in R}\psi^\dagger_\bfk\bfu_{\bfk\mathbb{M}}\psi_\mathbb{M}+H.c.$, where $\bfu_{\bfk i}$ denotes the tunneling rate between the substructures.

\subsection{Effective electron-electron interactions}
It should be noted that the coupling to molecular vibrations introduces correlations between the electrons, however, except for the temperature dependence the results discussed below do not depend on this choice of interactions. The interacting contribution can be replaced or complemented with direct electron-electron interactions of Coulomb type. This statement is substantiated by that
\begin{align}
\sum_i
	\Bigl(
		\Psi^	\dagger H_1Q_i\Psi+\omega_ib^\dagger_ib_i
	\Bigr)=&
	\sum_i
		\omega_iB^\dagger_iB_i
		-
		\frac{1}{\Omega_0}
		\Bigl(
			\Psi^	\dagger H_1\Psi
		\Bigr)^2
	,
\label{eq-effUmodel}
\end{align}
where $B_i=b_i^\dagger+\Psi^\dagger H_1\Psi$ is a shifted bosonic operator which can be integrated out from the model. Thereby, the physics is transformed into a purely fermionic model with effective electron-electron interactions -- second term in Eq. \eqref{eq-effUmodel}, where $\Psi=(\psi_1\ \psi_2\ \ldots\ \psi_\mathbb{M})^t$ and $1/\Omega_0=\sum_i1/\omega_i$. Since both repulsive and attractive interactions lead to correlated states in the molecular structure, the origin of the correlations is immaterial for the discussion here.

It is meaningful to consider the explicit structure of $\Hamil_I=-(\Psi^\dagger H_1\Psi)^2/\Omega_0$, which assumes the form
\begin{align}
\Hamil_1=&
	-\frac{1}{\Omega_0}
	\sum_{\stackrel{\scriptstyle mm'}{ss'}}
	\biggl(
		t_1^2n_{mm+s}n_{m'm'+s'}
		-
		i4t_1\lambda_1\bfv_m^{(s)}\cdot\bfs_{mm+2s}n_{nn+s'}
\nonumber\\&
		-
		4\lambda_1^2\bfv_m^{(s)}\cdot\bfs_{mm+2s}\bfs_{m'm'+2s'}\cdot\bfv_{m'}^{(s')}
	\biggr)
	,
\end{align}
where the notation $n_{mm+s}=\psi^\dagger_n\psi_{n+s}$ and $\bfs_{mm+2s}=\psi^\dagger_m\bfsigma\psi_{m+2s}/2$. Pursuing the discusssion in terms of the non-diagonal charge ($n_{mm+s}$) and spin ($\bfs_{mm+2s}$) is viable since the site index $m$ is not a good quantum number in a chain of coupled sites.

While the charge-charge interactions (first term) are independent of the molecular geometry, the spin-charge (second term) and spin-spin (third term) interactions strongly depend on the geometry via the curvature $\bfv_m^{(s)}$. For instance, a strictly linear chain, see Fig. \ref{fig-schematic} (a), has a trivial curvature ($\bfv_m^{(s)}\equiv0$) such that the spin interactions vanish. By contrast, spin interactions exist in structures with non-trivial curvature, see Fig. \ref{fig-schematic} (b), (c). It should also be noticed that the spin-spin interactions have the form of a dipole-dipole interaction, where the dipolar orientation is determined by the curvature $\bfv_m^{(s)}$.

It is clear that for non-interacting electrons, the expectation values $\av{\bfs_{mn}}\equiv0$ in the absence of external magnetic fields. The molecular states obtained through the Hamiltonian $\Hamil_\text{mol}$, either as given in Eq. \eqref{eq-mol} or in the form discussed here, however, are correlated and, therefore, vanishing moments $\av{\bfs_{mn}}$ is not mandatory. It follows that spin-configurations may contribute to the total energy. Hence, the following discussion pertains to structures in which $\bfv_m^{(s)}\neq0$ for at least one $m$.

By rewriting the spin-operators as $\bfs_{mn}=\av{\bfs_{mn}}+\bfs_{mn}-\av{\bfs_{mn}}$, the product $\bfs_{nm+2s}\bfs_{mn+2s'}$ can be approximated by the mean-field expression
\begin{align}
\bfs_{mm+2s}\bfs_{nn+2s'}\approx&
	-
	\av{\bfs_{mm+2s}}
	\av{\bfs_{nn+2s'}}
\nonumber\\&
	+
	\av{\bfs_{mm+2s}}
	\bfs_{nn+2s'}
	+
	\bfs_{mm+2s}
	\av{\bfs_{nn+2s'}}
	,
\end{align}
assuming that $(\bfs_{mm+2s}-\av{\bfs_{mm+2s}})(\bfs_{nn+2s'}-\av{\bfs_{nn+2s'}})$ is negligible. The dipole-dipole interaction is, then, approximately
\begin{align}
\bfv_m^{(s)}\cdot\bfs_{mm+2s}&\bfs_{nn+2s'}\cdot\bfv_{n}^{(s')}
\approx
	\bfv_m^{(s)}\cdot
	\Bigl(
		\av{\bfs_{mm+2s}}
		\av{\bfs_{mm+2s'}}
\nonumber\\&
		-
		\av{\bfs_{mm+2s}}
		\bfs_{nn+2s'}
		-
		\bfs_{mm+2s}
		\av{\bfs_{nn+2s'}}
	\Bigr)
	\cdot\bfv_{n}^{(s')}
	.
\end{align}

For the purpose of making further analytical progress and without loss of generality, all terms in the Hamiltonian proportional to $t_0$ and $t_1$ can be discarded, since the processes associated with these mechanism are spin-independent and can neither induce nor destroy the spin-structure in the model. For a chain with four sites, the effective single electron Hamiltonian for the molecular electronic structure can be written
\begin{align}
\Hamil_\text{MF}=&
	\sum_{m=1,2}
		\Psi_m^\dagger
		\begin{pmatrix}
			\dote{m} & ih_\perp\bfv_m^{(+)}\cdot\bfsigma \\
			(-i)h_\perp\bfv_m^{(+)}\cdot\bfsigma & \dote{m+2}
		\end{pmatrix}
		\Psi_m
	,
\label{eq-HMF}
\end{align}
where the spinor $\Psi_m=(\psi_m\ \psi_{m+2})^t$, $h_\perp(M_0)=\lambda_0-i2\lambda_1^2M_0/\Omega_0$, and $M_0=\sum_{ms}\bfv_m^{(s)}\cdot\av{\bfs_{mm+2s}}$. It can be readily seen that $M_0$ is odd under time-reversal operation, as it should by inheriting its properties from the expectation value $\av{\bfs_{mm+2s}}$. Hence, the spins interact between next-nearest sites which, on the one hand is not surprising given that the spin-orbit coupling, both the direct and the vibrationally assisted, is of next nearest-site nature, while on the other hand simplifies the mathematical construction in addressing the spin structure.

Assuming that $\dote{m}=\dote{0}$, for all $m=1,...,4$, the spectra of the two matrices are given by
\begin{align}
E_{m\pm}(M_0)=&
	\dote{0}
	\pm
		|\bfv_m^{(+)}|
		h_\perp(M_0)
	,\
	m=1,2
	.
\label{eq-Empm}
\end{align}
Each of the energies $E_{m\pm}$ are two-fold degenerate as a consequence of that the matrices in the Hamiltonian in Eq. \eqref{eq-HMF} both are four dimensional. In the full system, comprising four sites, these energies are, hence, four-fold degenerate.

In can be noticed that, the direct spin-orbit coupling opens a gap of width $2\lambda_0$, as expected for a spin-orbit coupled system, while the interactions give rise to a life-time broadening. Hence, under the assumption that the expectation values $\av{\bfs_{mm+2s}}$ are real valued, this imaginary part vanishes identically since the off-diagonal spins, or coherences, must fulfill the condition $\av{\bfs_{m+2m}}=\av{\bfs_{mm+2}}^*$. Therefore, in the closed shell system, the influence of the interactions is suppressed and has only a minor impact on the single-electron spectrum.

By contrast, should the coherences $\av{\bfs_{mm+2s}}$ not be real valued, a gap opens in the spectrum which indicates a possibility of an spontaneous time-reversal symmetry breaking. This is a direct consequence of the requirement $\av{\bfs_{m+2m}}=\av{\bfs_{mm+2}}^*$, since then
\begin{align}
M_0=&
	\sum_m\bfv_m^{(s)}\cdot\av{\bfs_{mm+2s}}
	=
	\bfv_1^{(+)}\cdot\av{\bfs_{13}-\bfs_{31}}+\bfv_2^{(+)}\cdot\av{\bfs_{24}-\bfs_{42}}
\nonumber\\=&
	i2\im
	\Bigl(
		\bfv_1^{(+)}\cdot\av{\bfs_{13}}
		+
		\bfv_2^{(+)}\cdot\av{\bfs_{24}}
	\Bigr)
	.
\end{align}
Hence, the spectrum for the four site is given by
\begin{align}
E_{m\pm}=&
	\dote{0}
	\pm
	\Biggl(
		\lambda_0
		+
		4\frac{\lambda_1^2}{\Omega_0}
		\im
		\Bigl(
			\bfv_1^{(+)}\cdot\av{\bfs_{13}}
			+
			\bfv_2^{(+)}\cdot\av{\bfs_{24}}
		\Bigr)
	\Biggr)
	|\bfv_m^{(+)}|
	,
\end{align}
for $m=1,2$, that is, a non-vanishing spin texture defined by $\av{\bfs_{mm+2}}$ modifies the spectral gap.

From the above discussion it can be understood that the energy of the four-electron state is $2(E_{1-}+E_{2-})=4\dote{0}-2h_\perp(M_0)(|\bfv_1^{(+)}|+|\bfv_2^{(+)}|)$, due to the two-fold degeneracies. Hence, non-vanishing spin correlations, manifest in $\av{\bfs_{mm+2}}\neq0$, lowers the total energy, underlining that definite spin-configurations are favored over random paramagnetic states.

Using electron reservoircontour ordered Green functions \cite{Kadanoff&Baym,haug2008quantum,NENP2010}, it can be shown that the coherences $\av{\bfs_{mm+2s}}$ are non-vanishing in the open system that follows by coupling the molecule to the electron reservoir. To lowest order approximation the off-diagonal spin expectation values are given by
\begin{subequations}
\label{eq-scoh}
\begin{align}
\av{\bfs_{13}-\bfs_{31}}
	\approx&
	(-i)2\lambda_0\bfv_1^{(+)}\Gamma^LI_L
	,
\\
\av{\bfs_{24}-\bfs_{42}}\approx&
	(-i)2\lambda_0\bfv_2^{(+)}\Gamma^RI_R
	,
\end{align}
\end{subequations}
where $\Gamma^{L(R)}$ is the energy related to the coupling to the reservoir $L(R)$ and where the integral $I_{L(R)}=-2\int\im g^r_{1(4)}(\omega)\re g_{3(2)}^{r}(\omega)f_{L(R)}(\omega)d\omega/2\pi$ reflects the occupied electronic density at the sites involved in the corresponding spin distributions. Here, $g_m^{r/a}$ denotes the retarded/advanced on-site Green function and $f_{L(R)}(\omega)$ is the Fermi-Dirac distribution function for the reservoir $L(R)$.

With the expressions in Eq. \eqref{eq-scoh}, the molecular spectrum in Eq. \eqref{eq-Empm} becomes
\begin{align}
E_{m\pm}=&
	\dote{0}
	\pm
	\lambda_0
	\Biggl(
		1
		-
		8\frac{\lambda_1^2}{\Omega_0}
		\re
		\Bigl(
			|\bfv_1^{(+)}|^2\Gamma^LI_L
			+
			|\bfv_2^{(+)}|^2\Gamma^RI_R
		\Bigr)
	\Biggr)
	|\bfv_m^{(+)}|
	.
\end{align}

The first thing to notice is that the energy does not depend on the curvature of the structure, merely its existence. This is an important observations since, in this model, the total energies for two enantiomers necessarily have to be equal. Simultaneously, the curvature $\bfv_m^{(s)}$ defines the orientation of the spin-coherence, see Eq. \eqref{eq-scoh}. Furthermore, the product $(-i)\bfv_m^{(s)}$ represents the angular momentum of the electrons transferring between next-nearest sites and is odd under time-reversal operation. It cannot be stressed enough that, this angular momentum exists as a latent quanlity in the closed shell structure and is suppressed by conservation of energy, momentum, and angular momentum. These conservation requirements are relaxed when the molecule is coupled to the electron reservoir, in close analogy with the discussion in Ref. \cite{JPhysChemLett.14.5119}.

Second, it is important to notice that the stabilizing energy, the gap $|E_{m+}-E_{m-}|$, for the spin-configuration scales with the coupling strengths $\Gamma^\chi$. Third, the integral $I_\chi$ provides a measure of the electron leakage between the molecule and the reservoir, hence, in a sense a measure of the openness of the molecule. The electronic density at site $m\in\{1,2,3,4\}$ is described by the retarded/advanced Green function $g^{r/a}_m(\omega)=(\omega-\dote{m}+i\Gamma_m/2)$, where $\Gamma_m$ denotes the life-time broadening due to hybridization and collisions. When the electronic energy $\dote{m}$ is far below (above) the chemical potential of the reservoir, that is, in the limit $(\dote{m}-\mu_{L/R})/\Gamma_m\rightarrow-\infty$ [$(\dote{m}-\mu_{L/R})/\Gamma_m\rightarrow\infty$], the corresponding occupation number $\av{n_m}\rightarrow1$ [$\av{n_m}\rightarrow0$]. Then, the molecule may again be considered as a closed shell structure since the influence from the reservoir is negligible. However, whenever the ratio $(\dote{m}-\mu_{L/R})/\Gamma_m$ is finite, the occupation number $\av{n_m}\in(0,1)$ which is natural when the electron density leaks between the molecule and reservoir. Although the integral $I_\chi$ does not strictly equal the occupation number of any site $m$, the same logic can be applied to this quantity, suggesting that $|I_\chi|>0$ whenever $(\dote{m}-\mu_{L/R})/\Gamma_m$ is finite. The conclusion from these two observations is that the coupling to the reservoir, which opens the molecule for leakage, facilitates the stabilization of one structurally determined spin-configuration.

It should be emphasized that the lower the molecular vibration energy $\Omega_0$, the more energetically favorable is a non-trivial spin texture. This is in direct analogy with the result from the Hubbard model with on-site Coulomb repulsion \cite{Fradkin2013}.


\subsection{Molecular electronic structure}
For the remainder of this article, the discussion will be centered around the original model in Eq. \eqref{eq-mol}. Given this model, the electronic structure of the molecule is calculated using the contour ordered (non-equilibrium) Green function formalism \cite{Kadanoff&Baym,haug2008quantum,NENP2010}. Following this procedure guarantees that the resulting Green function maintains the required time-reversal symmetries.

Using non-equilibrium Green functions allows a discussion on general grounds. For example, the local charge $\av{n_m}$, spin $\av{\bfs_m}$, and density of electron states $\rho_m$  are obtained by identifying $\av{n_m}=(-i){\rm sp}\int\bfG^<_{mm}(\omega)d\omega/2\pi$, $\av{\bfs_m}=(-i){\rm sp}\bfsigma\int\bfG^<_{mm}(\omega)d\omega/4\pi$, and $\rho_m(\omega)=i{\rm sp}[\bfG_{mm}^>(\omega)-\bfG_{mm}^<(\omega)]$, where ${\rm sp}$ denotes the trace over spin 1/2 space. Here, $\bfG^<_{mn}(\omega)$ and $\bfG_{mm}^>(\omega)$ denotes the lesser and greater Green functions, respectively, which are proportional to the molecular density of occupied and unoccupied electrons states.

The model $\Hamil=\Hamil_\text{mol}+\sum_\chi\Hamil_\chi+\Hamil_T$ allows to define the molecular single electron Green function $\bfG=\av{\inner{\Psi}{\Psi^\dagger}}$ in the form of a Dyson equation $\bfG(z)=\bfg(z)+\bfg(z)\bfSigma(z)\bfG(z)$, where $\bfg$ denotes the unperturbed Green function. The Dyson equation implies that $\bfG^{</>}_{mn}=\bfG^r_{m\mu}\bfSigma^{</>}_{\mu\nu}\bfG^a_{\nu n}$, where summation over repeated indices is understood and where $\bfG^{r/a}$ denotes the retarded/advanced Green functions. Furthermore, the Dyson equation also permits an order-by-order expansion of the Green functions of the form $\bfG^{r/a}=\bfg^{r/a}+\bfg^{r/a}\bfSigma^{r/a}\bfG^{r/a}$, which enables a in-depth analysis of the viable processes underlying the electronic structure.

The lesser/greater self-energy is defined by \cite{PhysRevB.102.235416} $\bfSigma^{</>}=H_1\Sigma^{</>}_\text{ph}H_1$ and which is site-resolves as
\begin{align}
\bfSigma_{mn}^{</>}(\omega)=&
	\sum_{ss'=\pm1}
	\sum_\mu
		\Bigl(
			-t_1\delta_{\mu m+s}
			+
			i\lambda_1\bfv_m^{(s)}\cdot\bfsigma\delta_{\mu m+2s}
		\Bigr)
\nonumber\\&\times
		\bfB_\mu^{</>}(\omega)
		\Bigl(
			-t_1\delta_{n\mu+s'}
			+
			i\lambda_1\bfv_\mu^{(s')}\cdot\bfsigma\delta_{n\mu+2s'}
		\Bigr)
	,
\label{eq-Sigmaless}
\end{align}
where
\begin{align}
\bfB_m^{</>}(\omega)\equiv&
	\Bigl(\Sigma^{</>}_\text{ph}(\omega)\Big)_m
	=
	i
	\sum_{ij}
	\int
		\bfg^{</>}_m(\omega-\dote{})D_{ij}^{</>}(\dote{})
	\frac{d\dote{}}{2\pi}
	,
\label{eq-Sigmavibless}
\end{align}
denotes the vibrational self-energy. In this expression $D_{ij}(z)=2\delta_{ij}\omega_i/(z^2-\widetilde\omega_i)$ is the vibrational Green function, with renormalized vibrational energy $\widetilde\omega_i=\omega_i\sqrt{1-(6\Phi/\omega_i)^2\coth(\beta\omega_i/2)}$ corresponding to the intra-mode restricted Hartree approximation of $\Hamil_\text{vib}$. It is, furthermore, understood that contribution with labels outside the range $1\leq m\leq\mathbb{M}$ identically vanish.

The retarded/advanced self-energy is defined by \cite{PhysRevB.102.235416} $\bfSigma^{r/a}=H_0+H_1\Sigma^{r/a}_\text{ph}H_1$, giving the site resolved expression
\begin{align}
\bfSigma_{mn}^{r/a}=&
	\sum_{s=\pm1}
	\Biggl(
		\Bigl(
			-t_0\delta_{nm+s}
			+
			i\lambda_0\bfv_m^{(s)}\cdot\bfsigma\delta_{nm+2s}
		\Bigr)
	+
	\sum_{s'\mu}
		\Bigl(
			-t_1\delta_{\mu m+s}
\nonumber\\&
			+
			i\lambda_1\bfv_m^{(s)}\cdot\bfsigma\delta_{\mu m+2s}
		\Bigr)
		\bfB_\mu^{r/a}(\omega)
		\Bigl(
			-t_1\delta_{n\mu+s'}
\nonumber\\&
			+
			i\lambda_1\bfv_\mu^{(s')}\cdot\bfsigma\delta_{n\mu+2s'}
		\Bigr)
	\Biggr)
	,
\label{eq-Sigmaret}
\end{align}
where
\begin{align}
\bfB_m^{r/a}(\omega)=&
	i
	\int
		\frac{
			\bfB_m^>(\dote{})
			-
			\bfB_m^<(\dote{})
		}
		{\omega-\dote{}\pm i\delta}
	\frac{d\dote{}}{2\pi}
	.
\label{eq-Sigmavibret}
\end{align}

Henceforth, the attention is put to the influence from the reservoir on the molecular structure. Hence, corrections that do not directly involve couplings to the edge sites will be discarded for the sake transparency. The principle of the calculations follow the order-by-order expansion of the lesser Green function
\begin{align}
\bfG_{mm}^<=&
	\bfg_m^<
	+
	\bfg_1^r
	\Bigl(
		\bfSigma_{mm}^<
		+
		\bfSigma_{mn}^r\bfg_n^r\bfSigma_{nm}^<
		+
		\bfSigma_{mn}^<\bfg_n^a\bfSigma_{nm}^a
		+
		\cdots
	\Bigr)
	\bfg_1^a
	.
\label{eq-Glesser}
\end{align}
Here, $\bfg_m^<=if_m\bfg_m^r\bfGamma_m\bfg_m^a$ defines the density of occupied density of non-interacting electron states at site $m$, which essentially is provided by the Fermi-Dirac distribution function $f_m$ centered around the electron density $\bfg_m^r\bfGamma_m\bfg_m^a$. The distribution functions at the edge sites are determined by the chemical potential of the reservoirs, $f_{1/\mathbb{M}}=f_{L/R}$ with $\mu_{L/R}=\mu_0\pm eV/2$ in terms of the voltage bias $V$, whereas $\mu_m=\mu_0$ for $2\leq m\leq\mathbb{M}-1$, hence, writing $f_m=f_0$.

The first significant observation that should be made from the expansion of the Green function in Eq. \eqref{eq-Glesser} is that coupling between the molecule and the reservoir has significant consequences for the molecular electronic structure. That the edge sites are broadened by this coupling is natural and can be quantified by the coupling matrix $\bfGamma_m=\bfGamma^\chi=-2\im\bfSigma^r_\chi$, such that $\bfg_m^{r/a}=(\omega-\dote{0}\pm i\bfGamma^\chi/2)^{-1}$. Here, the self-energy $\bfSigma_\chi^r=\sum_{\bfk\in\chi}\bfu_{\bfk i}^\dagger g_\bfk^r\bfu_{\bfk i}$ couples the molecule with electronic structure of the reservoir via the Green function $g_\bfk^r$. Hence, the electronic structure at the edge sites, 1 and $\mathbb{M}$, are directly influenced by the electronic and magnetic properties of the left and right lead, respectively. This is determined by the coupling $\bfGamma^\chi=\Gamma^\chi(\sigma^0+\bfp_\chi\cdot\bfsigma)/2$, where $\bfp_\chi$, $p_\chi=|\bfp_\chi|\leq1$, is the spin-polarization in reservoir $\chi$.

However, the sites in the interior of the molecule are also broadened by the interfacing to the reservoir. To lowest order, these broadenings are for the second and third sites from the reservoirs formally given by $\bfGamma_2=-2\im(\bfSigma_{22}^r+\bfSigma_{21}^r\bfg_1^r\bfSigma_{12}^r)$ and $\bfGamma_3=-2\im(\bfSigma_{33}^r+\bfSigma_{31}^r\bfg_1^r\bfSigma_{13}^r)$. The expression in Eq. \eqref{eq-Sigmaret} leads to
\begin{subequations}
\label{eq-Gamma23}
\begin{align}
\bfGamma_2(\omega)=&
	-2
	\biggl(
		t_0^2
		+
		t_1^2\Coth\frac{\beta\tilde\omega_0}{2}
	\biggr)
	\im~\bfg_1^r(\omega)
	,
\label{eq-Gamma2}
\\
\bfGamma_3(\omega)=&
	-2
	\biggl(
		\lambda_0^2
		+
		\lambda_1^2\Coth\frac{\beta\tilde\omega_0}{2}
	\biggr)
	|\bfv_1^{(+)}|^2\im~\bfg_1^r(\omega)
	.
\label{eq-Gamma3}
\end{align}
\end{subequations}
where it is assumed that the vibrational energies $\tilde\omega_i$ are small enough to allow the approximations $\bfB_m^{</>}(\omega)\approx(\pm i)2\pi f_0(\omega)\delta(\omega-\dote{0})\Coth(\beta\tilde\omega_0/2)$ for $2\leq m\leq\mathbb{M}-1$ and $\bfB_m^{</>}(\omega)\approx\bfg_m^{</>}(\omega)\Coth(\beta\tilde\omega_0/2)$ for $m=1,\mathbb{M}$, where $\Coth(\beta\tilde\omega_0/2)\equiv\sum_i\coth(\beta\tilde\omega_i/2)$. Therefore, e.g., $\bfB_1^r\approx\bfg_1^r\Coth(\beta\tilde\omega_0/2)$. Analogous expressions can be written for the sites $\mathbb{M}-1$ and $\mathbb{M}-2$ near the right reservoir.

The expressions in Eq. \eqref{eq-Gamma23} explicitly shows that the level broadening in the interior of the molecule is (\emph{i}) facilitated by the electron correlations and (\emph{ii}) acquires a spin-dependence from the reservoir whenever it is ferromagnetic. While these features exist also for non-interacting electrons, the electron correlations enhance them deeper inside the molecular structure. Indeed, by electron correlations, the broadening and spin-polarization propagates further into the molecule. The proximity effect that spin-polarizes the edge sites is extended deeper inside the structure by the interactions. Hence, following this logic the Green function for site $m$ are set to the spin-dependent form $\bfg^{r/a}_m(\omega)=1/(\omega-\dote{m}\pm i\bfGamma_m/2)^{-1}$, with the broadening defined according to the above discussion.

The effects of electron correlations is investigated using the forms of the self-energies in Eqs. \eqref{eq-Sigmaless} and \eqref{eq-Sigmaret}, however, suppressing the induced spin-dependence of $\bfGamma_m$, $m=2,3$, c.f., Eq. \eqref{eq-Gamma23}, for transparency and without loss of generality. It is, then, readily observed that the (un-normalized) charge and spin densities at site $m=1,\mathbb{M}$ are, up to second order in the self-energy, given by
\begin{subequations}
\label{eq-ns1}
\begin{align}
\av{n_m}\approx&
	\av{n_m}^{(0)}
	-
	\frac{128}{\Gamma^\chi}
	\biggl(
		\frac{t_1^2}{\lambda_0|\bfv_m|^2}
		-
		\frac{t_1\lambda_1}{t_0}
	\biggr)
	\frac{\bfp_\chi\cdot\bfv_m}{(1-p_\chi^2)^2}
	\Coth\frac{\beta\tilde\omega_0}{2}
	,
\label{eq-n1}
\\
 \av{\bfs_m}\approx&
 	\av{\bfs_m}^{(0)}
	+
	\frac{32}{\Gamma^\chi}
	\biggl(
		\frac{t_1^2}{\lambda_0|\bfv_m|^2}
		-
		\frac{t_1\lambda_1}{t_0}
	\biggr)
	\biggl(
		\frac{\bfv_m}{1-p_\chi^2}
\nonumber\\&
		+
		\frac{\bfp_\chi\cdot\bfv_m}{(1-p_\chi^2)^2}
		\bfp_\chi
	\biggr)
	\Coth\frac{\beta\tilde\omega_0}{2}
	,
\label{eq-s1}
\end{align}
 \end{subequations}
where $\bfv_1\equiv\bfv_1^{(+)}$ and $\bfv_\mathbb{M}\equiv\bfv_\mathbb{M}^{(-)}$, whereas $\av{n_m}^{(0)}=(-i){\rm sp}\int\bfg_m^<(\omega)d\omega/2\pi$ and $\av{\bfs_m}^{(0)}=(-i){\rm sp}\bfsigma\int\bfg_m^<(\omega)d\omega/4\pi$ define the non-interacting/uncorrelated charge and spin densities, respectively, for the molecule interfaced with the reservoirs.

The significance of the results in Eq. \eqref{eq-ns1} is important to stress since they explicitly demonstrate that (\emph{i}) a chiral molecule acquire a specific spin-configuration upon interfacing with a reservoir, and (\emph{ii}) the charge density of the molecule depends directly on the spin-polarization of a ferromagnetic reservoir. The first point becomes clear by setting the spin-polarization parameter $\bfp_\chi=0$. Then, the second term to the right in Eq. \eqref{eq-s1} is directly proportional to $\bfv_m$ and, therefore, only vanishes for strictly linear structures, c.f., Fig. \ref{fig-schematic}. Hence, while an achiral zig-zag structure may acquire a transverse spin-configuration, a chiral also acquires a longitudinal one. Moreover, for  a regular helical structure, the longitudinal components of the curvature vectors $\bfv_1$ an $\bfv_\mathbb{M}$ have opposite signs. This leads to that $\av{s_1^z}=-\av{s_\mathbb{M}^z}$, since $\av{\bfs_m}^{(0)}=0$. As a result, the molecule/reservoir heterostructure is in a spontaneously time-reversal broken state in which the molecular spin-configuration is governed by the chirality. The Kramers doublet is, thereby, constituted by the two enantiomers as they are degenerate with respect to all properties except their chiralities and, hence, their spin-configurations.

The second point is explicit in the second term to the right in Eq. \eqref{eq-n1}, proportional to $\bfp_\chi\cdot\bfv_m$. Thus, the charge density varies with the spin-polarization in the reservoir. Such a response is a direct sign that no linear response regime as function of the external spin-polarization can be established, since the properties of the system change as function of the parameter $\bfp_\chi$. For a given curvature $\bfv_m$, the charge density $\av{n_m}$ is larger or smaller than the corresponding non-interacting value $\av{n_m}^{(0)}$ depending on $\bfp_\chi$ through the sign of $\bfp_\chi\cdot\bfv_m$. Since the charge current through the molecular junction, when biased with a voltage, directly depends on molecular charge density, the asymmetric response of the charge density as function of $\bfp_\chi$ provides a sufficient condition for the system to yield a charge current that depends linearly with $\bfp_\chi$.

The implication of these two result is that the Onsager reciprocity theorem does not apply for a chiral molecule interfaced with a reservoir, since the system is in a time-reversal broken state and no linear response regime can be established with respect to the external spin-polarization.

\subsection{Numerical examples}
The numerical calculations presented here are based on non-equilibrium Green functions and the matrix inversion $\bfG^{r/a}(\omega)=(\omega-E-\bfSigma^{r/a}(\omega))^{-1}$, with $\bfSigma^{r/a}$ as defined in Eq. \eqref{eq-Sigmaret}, and subsequently calculating $\bfG^{</>}(\omega)=\bfG^r(\omega)\sum_\chi\bfSigma^{</>}(\omega)\bfG^a(\omega)$. In this way, the occupied and unoccupied electronic density is calculated in presence of the reservoir and including the electron correlations.

\begin{figure}[t]
\begin{center}
\includegraphics[width=\columnwidth]{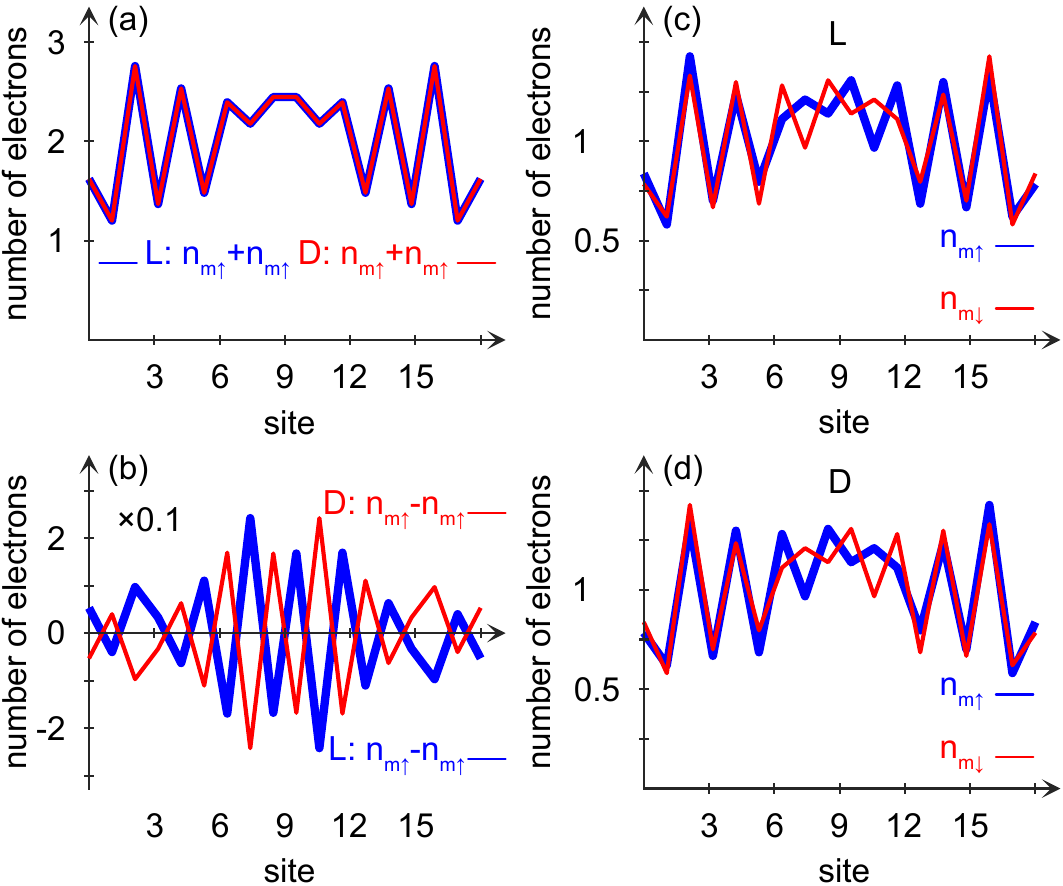}
\end{center}
\caption{Site resolved (a) charge and (b) spin distributions for helical $3\times6$ sites (blue) $L$ and (red) $D$ enantiomers mounted between non-magnetic reservoirs. Corresponding site and spin resolved charge distributions for the (c) $L$ and (d) $D$ enantiomers, respectively, with (blue) spin $\up$ and (red) spin $\down$ projections. Here, $\lambda_0=1/50$, $t_1=1/20$, $\lambda_1=1/500$, $\Gamma_0=1$, $\dote{m}-\mu_0=-10$, $\omega_0=1/50$, and $\Phi=3/500$ in units of $t_0=0.1$ eV, calculated for $T=300$ K.}
\label{fig-enantiomers}
\end{figure}

The plots in Fig. \ref{fig-enantiomers} show an example of the site resolved charge and spin distributions for helical $3\times6$ $L$ and $D$ enantiomers mounted between non-magnetic reservoirs. The molecular sites are distributed according to the coordinates $\bfr_m=(a\cos\varphi_m,a\sin\varphi_m,c_m)$, where $\varphi_m=2\pi(m-1)/(\mathbb{M}-1)$ and $c_m=c(m-1)/(\mathbb{M}-1)$ in terms of the radius $a$ and length $c$. The charge is distributed equally in the two enantiomers, as expected, Fig. \ref{fig-enantiomers} (a), whereas the spin distributions, Fig. \ref{fig-enantiomers} (b)--(d), are mirror images of one another as expected from Eq. \eqref{eq-s1}. The non-trivial spin distribution functions as a coercivity with respect to an external magnetization $\bfp_\chi||\hat{\bf z}$ such that the magnetization either confluent with or counteracting the spin polarization of the edge site, which is also explicit in Eq. \eqref{eq-s1}.

\begin{figure}[t]
\begin{center}
\includegraphics[width=\columnwidth]{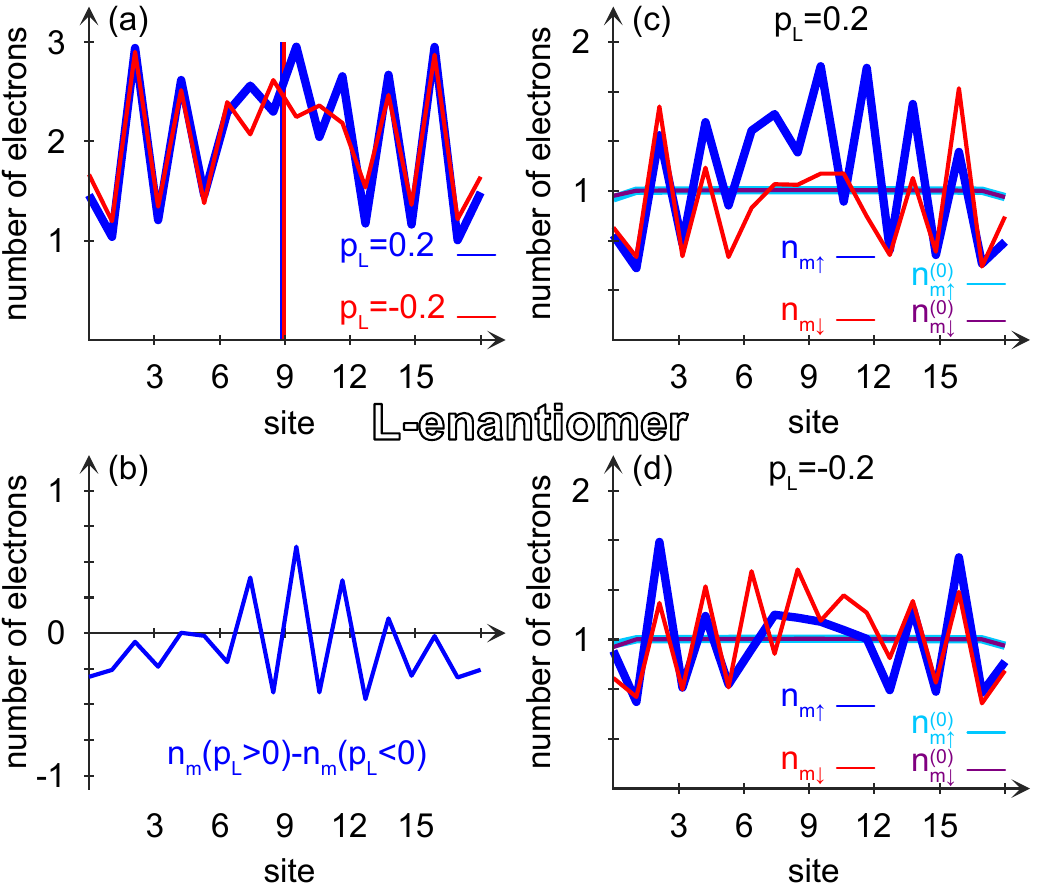}
\end{center}
\caption{(a) Site resolved charge distribution for a helical $3\times6$ sites $L$ enantiomer mounted between a ferromagnetic ($\bfp_L=\pm0.2\hat{\bf z}$) and a non-magnetic ($\bfp_R=0$) reservoirs for (blue) $\bfp_L=0.2\hat{\bf z}$ and (red) $\bfp_L=-0.2\hat{\bf z}$. Vertical lines signify the center of the charge polarization for the two conditions. (b) Difference between the charge distributions for $\bfp_L=\pm0.2\hat{\bf z}$. (c), (d) site and spin resolved charge distributions for (c) $\bfp_L=0.2\hat{\bf z}$ and (d) $\bfp_L=-0.2\hat{\bf z}$, with (blue) spin $\up$ and (red) spin $\down$ projections. Corresponding unperturbed charge distributions $n_{m\sigma}^{(0)}$ for (cyan) $\sigma=\up$ and (purple) $\sigma=\down$ are included as reference. Parameters are as in Fig. \ref{fig-enantiomers}.}
\label{fig-Lenantiomer}
\end{figure}

\begin{figure}[t]
\begin{center}
\includegraphics[width=\columnwidth]{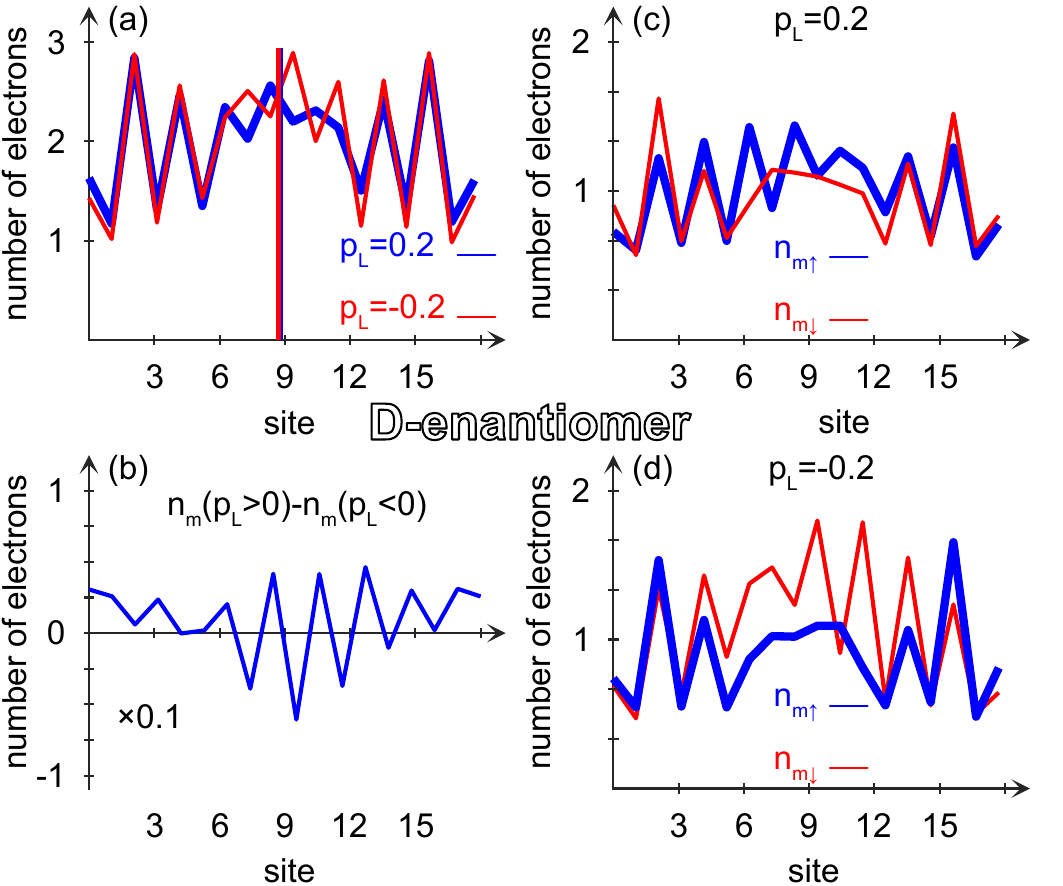}
\end{center}
\caption{(a) Site resolved charge distribution for a helical $3\times6$ sites $D$ enantiomer mounted between a ferromagnetic ($\bfp_L=\pm0.2\hat{\bf z}$) and a non-magnetic ($\bfp_R=0$) reservoirs for (blue) $\bfp_L=0.2\hat{\bf z}$ and (red) $\bfp_L=-0.2\hat{\bf z}$. (b) Difference between the charge distributions for $\bfp_L=\pm0.2\hat{\bf z}$. (c), (d) site and spin resolved charge distributions for (c) $\bfp_L=0.2\hat{\bf z}$ and (d) $\bfp_L=-0.2\hat{\bf z}$, with (blue) spin $\up$ and (red) spin $\down$ projections. Parameters are as in Fig. \ref{fig-enantiomers}.}
\label{fig-Denantiomer}
\end{figure}

The asymmetric influence of a ferromagnetic reservoir, $\bfp_L=\pm0.2\hat{\bf z}$, $\bfp_R=0$, is illustrated in Figs. \ref{fig-Lenantiomer} and \ref{fig-Denantiomer}, which show the site and spin resolved charge distributions for the $L$ and $D$ enantiomers, respectively. Especially the asymmetric influence on the spin distributions by switching between $\bfp_L=\pm0.2\hat{\bf z}$ is clearly shown in Figs. \ref{fig-Lenantiomer} (c), (d), and \ref{fig-Denantiomer} (c), (d). A direct comparison between in Figs. \ref{fig-Lenantiomer} and \ref{fig-Denantiomer} also demonstrate that the results are mirror images of one another, as should be expected.

As shown in Eq. \eqref{eq-n1}, the charge distribution is expected to depend on the magnetization of the reservoirs, which is corroborated by the numerical results in Figs. \ref{fig-Lenantiomer} (a), (b), and \ref{fig-Denantiomer} (a), (b), showing the site resolved (a) total charge for $\bfp_L=\pm0.2\hat{\bf z}$ and (b) corresponding difference between these. From these plots one can deduce that there is more charge localized within the $L$ ($D$) enantiomer for $\bfp_L=0.2\hat{\bf z}$ ($\bfp_L=-0.2\hat{\bf z}$). A direct measure of the asymmetric response is the charge polarization $P_z=2\sum_m(c_m-\av{c})\av{n_m}/c\mathbb{M}$, where $\av{c}=\sum_mc_m/\mathbb{M}$, shows that $P_z(\bfp_L>0)<P_z(\bfp_L<0)$, as indicated by the vertical lines in Figs. \ref{fig-Lenantiomer} and \ref{fig-Denantiomer} (a). A deeper analysis reveal that the conditions leading to more localized charge also lead to that the charge is strongly spin-polarized, see Figs. \ref{fig-Lenantiomer} (c) and \ref{fig-Lenantiomer} (d). By contrast, the less localized charge is also less spin-polarized, see Figs. \ref{fig-Lenantiomer} (d) and \ref{fig-Lenantiomer} (c).

Moreover, as predicted by Eqs. \eqref{eq-Gamma23},  \eqref{eq-s1} for uncorrelated electrons, there is not expected to be any significant penetration depth of the ferromagnetism of the reservoir into the molecule. The calculations confirm this prediction, as can be seen in Fig. \ref{fig-Lenantiomer} (c), (d), where the site resolved unperturbed charge distributions $\av{n_{m\sigma}}^{(0)}$ are plotted for spin projections (cyan) $\sigma=\up$ and (purple) $\sigma=\down$. It should stressed that the unperturbed charge is homogeneously distributed and nearly spin degenerate except for the site adjacent to the left reservoir, where the nominal spin polarization is $\av{n_{m\up}}^{(0)}-\av{n_{m\down}}^{(0)}<0.02$. In contrast to correlated electrons, the total charge distribution $\av{n_{m\up}}^{(0)}+\av{n_{m\down}}^{(0)}$ for uncorrelated electrons is independent of the external magnetization. The reason is that uncorrelated electrons do not by themselves generate a magnetic anisotropy and are, hence, trivially polarizable in a paramagnetic sense.

\begin{figure}[b]
\begin{center}
\includegraphics[width=\columnwidth]{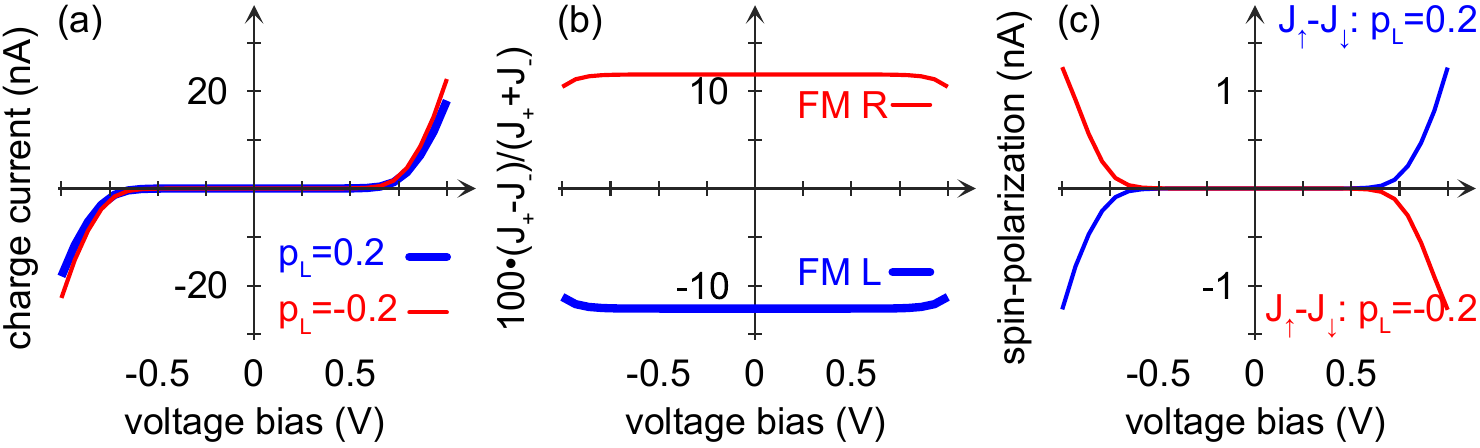}
\end{center}
\caption{(a) Charge currents for a $3\times6$ site $L$ enantiomer for (blue) $\bfp_L=0.2\hat{\bf z}$ and (red) $\bfp_L=-0.2\hat{\bf z}$, (b) the normalized magneto-current $100(J_+-J_-)/(J_++J_-)$ for (blue) $\bfp_L=\pm0.2\hat{\bf z}$, $\bfp_R=0$, and (red) $\bfp_L=0$, $\bfp_R=\pm0.2\hat{\bf z}$, and (c) spin-polarizations of the currents in panel (a). Parameters are as in Fig. \ref{fig-enantiomers}.}
\label{fig-currents}
\end{figure}

The stronger localization of charge for the $L$ ($D$) enantiomer at $\bfp_L=0.2\hat{\bf z}$ ($\bfp_L=-0.2\hat{\bf z}$) can be interpreted as a higher barrier for electron transport compared to the conditions at $\bfp_L=-0.2\hat{\bf z}$ ($\bfp_L=0.2\hat{\bf z}$). In other words, the more strongly the charge is localized within the molecular junction, the less charge current can flow under a voltage bias. This statement is confirmed by the charge currents calculated for the $L$ enantiomer for $\bfp_L=\pm0.2\hat{\bf z}$, as shown in Fig. \ref{fig-currents} (a). The currents are in both cases small in the voltage range between $|V|\lesssim7.5t_0/e=0.75$ V. Beyond this range, the current calculated for $\bfp_L=-0.2\hat{\bf z}$ (red) grows more rapidly that the one for $\bfp_L=0.2\hat{\bf z}$ (blue). Hence, the associated magneto-current $(J_+-J_-)/(J_++J_-)$ is negative, see Fig. \ref{fig-currents} (c). The magneto-current varies only weakly over a wide range of voltages and maintain the same sign for both positive and negative polarities, two observations that are often observed in experiments, see, e.g., Refs. \cite{NanoLett.11.4652,JPhysChemC.124.10776,NanoscaleHoriz.8.320,Small.20.2308233,JPhysChemLett.15.6605,ACSNano.19.17941}.

Switching the ferromagnetic reservoir from the left to the right side, that is, setting $\bfp_L=0$ and $\bfp_R=\pm0.2\hat{\bf z}$ leads to the opposite, positive, magneto-current, see Fig. \ref{fig-currents} (c). This result is in excellent agreement with the recent observations reported in \cite{ACSNano.19.17941}, in which the charge current was measured both as function of the sign of the magnetization and its spatial location in the junction. Using the framework presented here, this result can be explained in the following way. The charge distributions for the $L$ enantiomer configured with $\bfp_R=\pm0.2\hat{\bf z}$ can be obtained by reflecting the distributions for the $D$ enantiomer configured with $\bfp_L=\pm0.2\hat{\bf z}$ around the molecular midpoint (between sites 9 and 10). Hence, from a transport point of view, the $L$ ($D$) enantiomer acquires the magneto-current properties of the $D$ ($L$) enantiomer when the magnetic reservoir is moved from the left to the right side of the junction. This shift can also be predicted from molecular spin distributions acquired for non-magnetic reservoirs, see Fig. \ref{fig-enantiomers} (b), suggesting that the spin distribution for the $D$ enantiomer is obtained by reflecting the corresponding distribution for the $L$ enantiomer around the horizontal axis.

The plots in Fig. \ref{fig-currents} (c) show the spin-polarizations of the currents $J_\up-J_\down$ in Fig. \ref{fig-currents} (a) for (blue) $\bfp_L=0.2\hat{\bf z}$ and (red) $\bfp_L=-0.2\hat{\bf z}$. The non-vanishing spin-polarization of the charge current is in agreement with the detection of spin-polarized electrons in photo-electron spectroscopy \cite{Science.283.814,Science.331.894}. The spin-polarization of the current in the transport set-up is yet to be experimentally measured.

\begin{figure}[t]
\begin{center}
\includegraphics[width=\columnwidth]{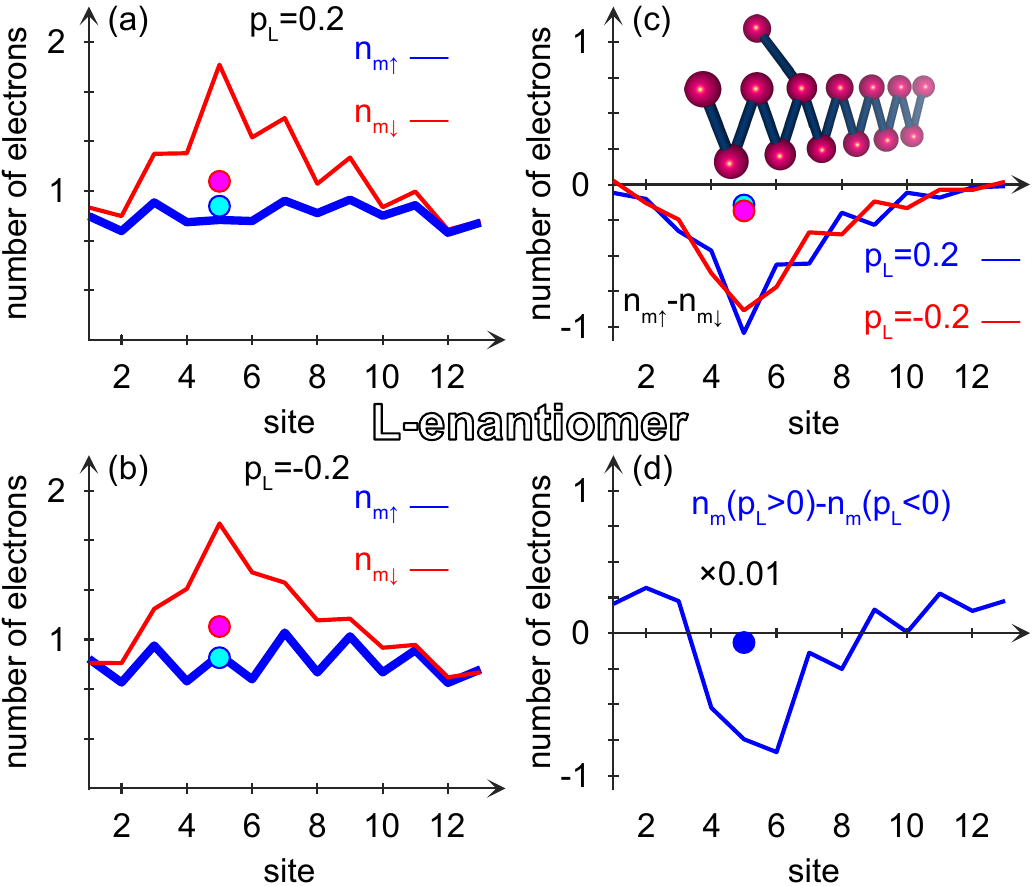}
\end{center}
\caption{
(a), (b) Site resolved non-equilibrium charge distributions showing (blue) $n_{m\up}$ and (red) $n_{m\down}$ for (a) $\bfp_L=0.2\hat{\bf z}$ and (b) $\bfp_L=-0.2\hat{\bf z}$, (c) corresponding spin-polarizations, and (d) difference in the total charge distributions in (a) and (b). Inset in (c) illustrates the geometry of the structure and the dots indicate the corresponding charge densities for the protruding site. Here, $V=0.5$ V, while other parameters are as in Fig. \ref{fig-enantiomers}.
}
\label{fig-CDzzw}
\end{figure}

Finally it is important to point out that the modeling presented in this paper is not limited to helical structures but it applicable to chiral structures in general. For this purpose, consider a zig-zag chain with a single site protruding from the plane defined by the zig-zag structure, see the inset of Fig. \ref{fig-CDzzw} (c). Comparing to the previous example, the only difference here is the geometry. It can be noticed that this chiral structure also responds anistropically to the magnetization in the reservoir, see Fig. \ref{fig-CDzzw} (a), (b), in which the site resolved non-equilibrium charge densities $n_{m\sigma}$ are plotted for the voltage bias $V=0.5$ V. The overall asymmetry between the spin channels is markedly different for the two magnetizations, which is also captured in the different spin-polarizations $n_{m\up}-n_{m\down}$ plotted in Fig. \ref{fig-CDzzw} (c), whereas the changes in the total charge distribution are less conspicuous, see Fig. \ref{fig-CDzzw} (d), which shows the difference in the charge distributions in panels (a) and (b). The dots indicate the corresponding values at the protruding site.

The largest effects on the spin configurations appear near the site (\#5) in the zig-zag chain which is coupled to the protruding site. This is not surprising since the chirality and, hence, the longitudinal curvature of the structure is centered around this site. Both the spin-polarizations and differences in the charge distributions are expected to the small in regions where the longitudinal curvature component is small, which is also corroborated by the example in Fig. \ref{fig-CDzzw}. 

\begin{figure}[t]
\begin{center}
\includegraphics[width=\columnwidth]{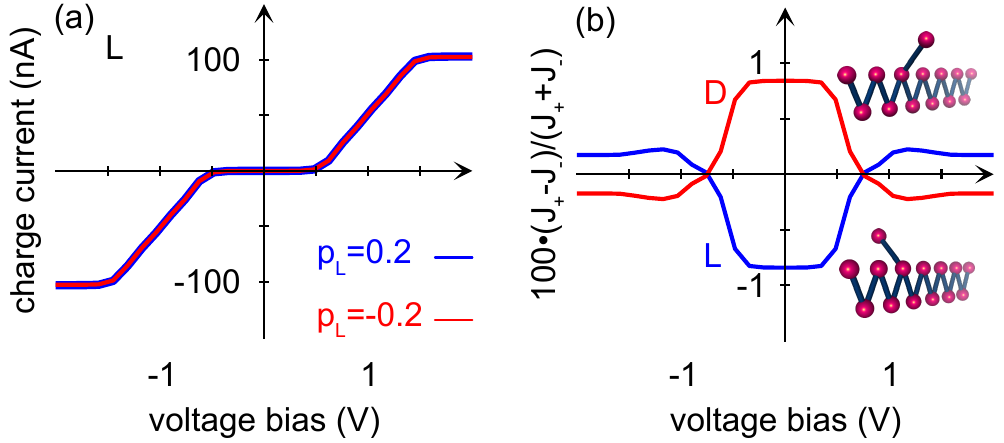}
\end{center}
\caption{
(a) Charge currents with (blue) $\bfp_L=0.2\hat{\bf z}$ and (red) $\bfp_L=-0.2\hat{\bf z}$, for the structure in Fig. \ref{fig-CDzzw}. (b) Normalized magneto-currents for (blue) the $L$- and (red) $D$-enantiomers shown in the insets.
Parameters are as in Fig. \ref{fig-enantiomers}.
}
\label{fig-IVzzw}
\end{figure}

Because of the small alterations between the charge distributions with varying external magnetization, the associated chirality induced spin selectivity effect is expected to be small, albeit non-vanishing. This is also verified numerically in Fig. \ref{fig-IVzzw} (a), where the charge currents are plotted as function of the voltage bias for (blue) $\bfp_L=0.2\hat{\bf z}$ and (red) $\bfp_L=-0.2\hat{\bf z}$. The currents a nearly degenerate, however, the normalized magneto-current, see Fig. \ref{fig-IVzzw} (b), is non-vanishing yet below 1 \%. The other enantiomer is also plotted (red) for completeness, showing the expected mirroring of the magneto-current around the horizontal axis.

\section{Discussion and Concluding Remarks}
The technical discussion in the foregoing section of the modeling and results is worthwhile to summarize from a phenomenological point of view. 

The theoretical picture emerging is founded on the basic fact that the molecular singlet state is a correlated state. This is crucial, since correlations put constraints electronic degrees of freedom when forming expectation values, especially concerning the spin.

Since the molecule is a closed shell structure when isolated from an electron reservoir, it is in a singlet state. While in average, the molecular spin moment vanishes, the molecular state can be regarded as fluctuating between different but equally probable spin-configurations. Moreover, the spin of an electron does not vary independently but depends exclusively on the spins of the other electrons. Suppression of the fluctuations leads to that the molecular spin-configuration stabilize and a local spin order emerges. As such, this order responds asymmetrically to an external magnetization.

In the present article, it is shown that the fluctuations between the different spin-configurations are suppressed for a chiral molecule interfaced with an electron reservoir. The molecular spin-configuration ultimately assumed is determined by chirality such that the two enantiomers of a chiral structure acquire enantiospecifc spin-configurations, in agreement with \cite{PhysRevB.108.235154}. The result that the molecule acquires a non-trivial enantiospecific spin-configuration is an emerging effect which necessitates the interfacing between the molecule with a reservoir.

The assumed spin-configuration has a direct connection with the molecular structure, or curvature, quantified in the vector $\bfv$. This structural quantity is a direct consequence of the electric field associated with electron confinement potential in the spin-orbit coupling, formally expressed as $i\bfv\cdot\bfsigma$, where $\bfsigma$ denotes the vector of Pauli matrices. From this expression it can be understood that any non-trivial curvature vector, $\bfv\neq0$, essentially pertains to the discussion herein.

Thus, as the chiral molecule is interfaced with an electron reservoir, the fluctuations between the spin-configurations are suppressed and the molecule is stabilized in one of these configurations. The assumed configuration is determined by the molecular chirality and can be understood to result from an effective spatially varying magnetic field provided by the product of the molecular spin-orbit coupling and the coupling $\Gamma$ to the electron reservoir. This coupling introduces a life-time broadening and is represented by the imaginary part of the electronic spectrum. Hence, the product $(-i\Gamma)(i\bfv\cdot\bfsigma)=\Gamma\bfv\cdot\bfsigma$ is real and provides a finite Zeeman-like contribution that enables the formation of a local magnetic moment.

The product $\Gamma\bfv\cdot\bfsigma$ is facilitated by effective molecular electron-electron interactions and does not arise for independent particles. The electron-electron interaction, which in the closed shell are conservative, dissipates energy and angular momentum into the reservoir. The losses effectively damps the molecular magnetic moments into one of the spin-configurations of the singlet state and the favored state is determined by the curvature vector $\bfv$.

A magnetized electron reservoir can be captured by setting the coupling $\bfGamma=\Gamma(\sigma^0+\bfp\cdot\bfsigma)$, where $\bfp$ parametrizes the spin-polarization. Then, the above product extends to (omitting a term with no consequence to the discussion)
\begin{align}
\bfGamma\bfv\cdot\bfsigma=&
	\Gamma\bfp\cdot\bfv
	+
	\Gamma\bfv\cdot\bfsigma
	,
\end{align}
which has two important features. The first term implies that the molecular charge depends linearly and enantiospecifically on the magnetization in the reservoir. Secondly, the spin-dependence, represented by the second term, includes a contribution to the spin-orbit coupling which also depends linearly on the magnetization in the reservoir. Hence, the molecule responds both asymmetrically and enantiospecifically to the external magnetization. This asymmetric response is ultimately associated with the distinct transport qualities referred to as the \emph{chirality induced spin selectivity} effect.

\section{Acknowledgements}
This work would not have been pursued was it not for Per Hedeg\aa rd and Jan van Ruitenbeek and their critical questions of the modeling I am repeatedly communicating with the community. I thank them for their patience. I am also indebted to Ron Naaman who keeps encouraging my investigations of chiral matter and spin, as well as Altug Sisman who asks the right critical questions whenever I think I have solved the problem. Finally, I thank the CISSors for keep attending the seminars and steadily providing new interesting results in this field. Olle Engkvist Stiftelse is acknowledged for financial support.

\bibliography{CISSref}

\end{document}